\documentclass{article}
\usepackage{arxiv}

\usepackage[utf8]{inputenc}
\usepackage[T1]{fontenc}
\usepackage[english]{babel}
\usepackage{amsmath,amsfonts,amssymb}
\usepackage{url}
\usepackage{booktabs}
\usepackage{nicefrac}
\usepackage{microtype}
\usepackage{graphicx}
\usepackage{xcolor}
\usepackage{array}
\usepackage{subcaption}
\usepackage{tcolorbox}
\tcbuselibrary{breakable}
\usepackage{comment}
\usepackage[hyperfootnotes=false]{hyperref}

\definecolor{VUB_blauw}{rgb}{0.1529,0.2667,0.5529}
\definecolor{mygrey}{HTML}{949494}

\newcommand{\customCor}[1]{\includegraphics[height=1em]{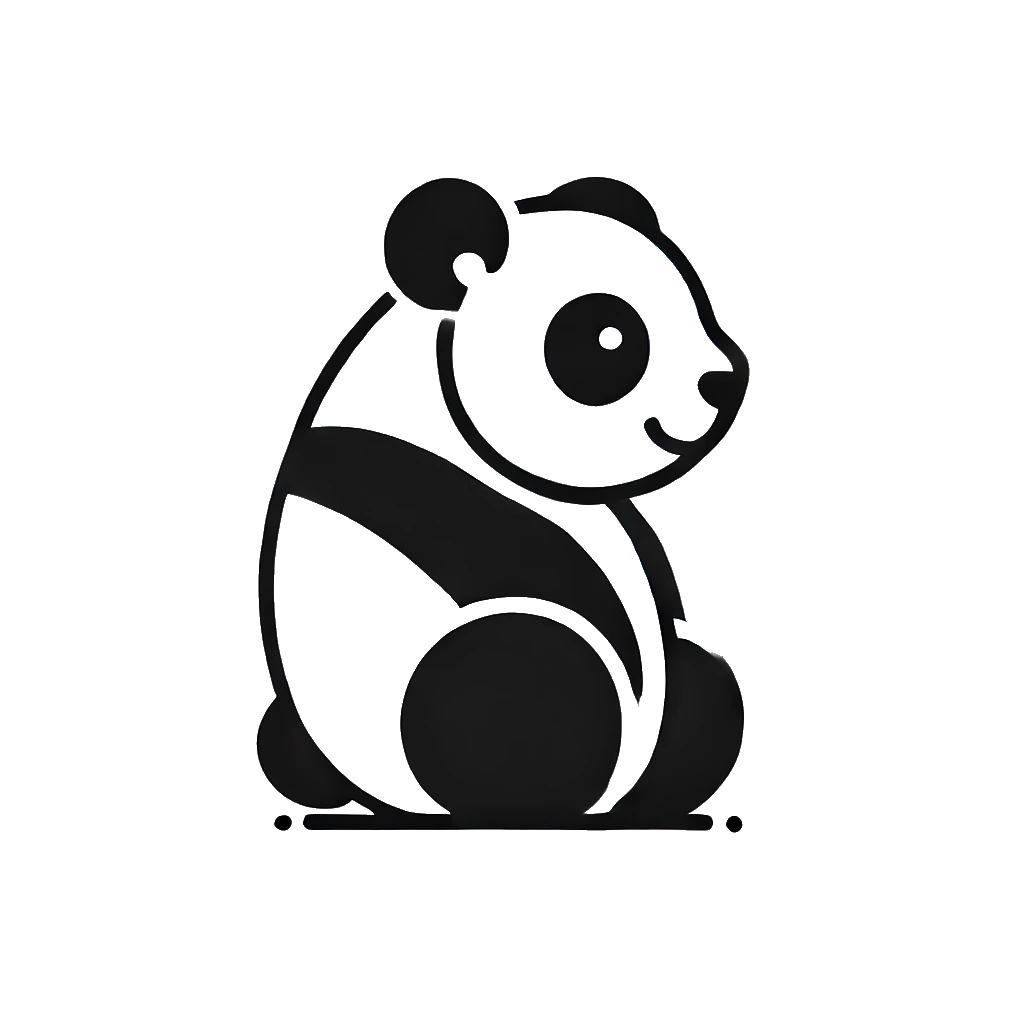} #1}
\hypersetup{
    colorlinks=true,
    citecolor=VUB_blauw,
    filecolor=VUB_blauw,
    linkcolor=VUB_blauw,
    urlcolor=VUB_blauw,
    breaklinks=true,
    pdftitle={Evaluating LLM-Generated Feedback on Handwritten Mathematics: Feedback Quality and Subsequent Exam Performance},
    pdfauthor={Arne Vanhoyweghen; Vincent Holst; Melika Mobini; Lukas Van de Voorde; Tibo Vanleke; Andres Algaba; Sam Verboven; Marie-Anne Guerry; Filip Van Droogenbroeck; Vincent Ginis}
}

\title{Evaluating LLM-Generated Feedback on Handwritten Mathematics: Feedback Quality and Subsequent Exam Performance}
\runningtitle{LLM-Generated Feedback on Handwritten Mathematics}

\author{
  Arne Vanhoyweghen\textsuperscript{1,2,\customCor{}} 
  \And
  Vincent Holst\textsuperscript{1,2}
  \And
  Melika Mobini\textsuperscript{1,2}
  \AND
  Lukas Van de Voorde\textsuperscript{1,2}
  \And
  Tibo Vanleke\textsuperscript{1,2}
  \And
  Andres Algaba\textsuperscript{1,2}
  \And
  Sam Verboven\textsuperscript{1}
  \AND
  Marie-Anne Guerry\textsuperscript{1}
  \And
  Filip Van Droogenbroeck\textsuperscript{1}
  \And
  Vincent Ginis\textsuperscript{1,2,3}
  \AND
  \normalfont\small\textsuperscript{1}Data Analytics Laboratory, Vrije Universiteit Brussel, Brussels, Belgium. \\
  \normalfont\small\textsuperscript{2}imec-SMIT, Vrije Universiteit Brussel, Brussels, Belgium.\\
  \normalfont\small\textsuperscript{3}School of Engineering and Applied Sciences, Harvard University, Cambridge, MA, USA.
}

\begin{document}
\raggedbottom
\maketitle
\begingroup
\renewcommand{\thefootnote}{}
\footnotetext{\includegraphics[height=1em]{panda2.png} Corresponding author: \href{mailto:arne.vanhoyweghen@vub.be}{arne.vanhoyweghen@vub.be}}
\endgroup
\thispagestyle{plain}

\begin{abstract}
We evaluate the quality of feedback and practice exercises generated by GPT-5.1 from handwritten university mathematics answers and their grades, and examine whether receiving this feedback improves subsequent exam performance. We use six weekly bonus tests in an English ($n=343$) and a Dutch ($n=236$) cohort. Tests~1 and~3 used a crossover feedback design; otherwise, all attendees received feedback. To evaluate the feedback, six reviewers assessed 157 messages from three English-cohort tasks using six criteria scored from 0 to 3. The practice exercises received the highest proportion of strong ratings (95.5\%), while diagnostic specificity (41.1\%) and clarity and tone (33.4\%) received fewer. Feedback for the lowest-scoring answers was rated as less actionable, suggesting that these students may need more targeted guidance from instructors. Comparisons of related exam questions showed mixed results, with no statistically significant differences after adjustment across four comparisons. Greater attendance and feedback exposure were associated with higher exam scores, but the contribution of feedback could not be separated from attendance and practice. Our findings show the potential of an integrated framework merging instructor guidance with automated feedback generation to provide personalised learning support at scale.
\end{abstract}

\keywords{Personalised Feedback \and Handwritten assessment \and Large language models}


\section{Introduction} 

\addcontentsline{toc}{section}{Introduction} 

A grade tells a student how well they performed, but it does not necessarily tell them what went wrong or how to improve. This creates an important gap between a course's evaluative and educational intent. Feedback is the most common way to close the gap between evaluation and learning,  by helping students identify their mistakes and understand what they need to work on \cite{hattie2007power,black1998assessment,shute2008focus,nicol2006formative,wisniewski2020power}. However, providing such feedback for handwritten answers is time-consuming and makes short feedback loops non-viable when the number of students increases. Previous work has used data from students' interactions with online learning activities to provide personalised feedback in large courses \cite{pardo2019using}. LLMs can assist human instructors by offering a way to quickly generate individual comments and practice exercises after an answer has been graded~\cite{meyer2026comparing, dai2024assessing,baral2024automated}.

We address two questions. First, can GPT-5.1 provide accurate feedback and practice exercises suited to the level demonstrated in a student's handwritten answer? Second, does receiving this feedback improve performance on related exam questions? A model might correctly explain how to solve an exercise but still provide little guidance to a student who does not know where to begin. It might also, because of hallucinations, offer advice on reasoning the student never used \cite{kasneci2023chatgpt,mollick2023assigning,steinbach2025llms}. Additionally, LLMs might provide useful feedback, but the feedback might fail to engage students to use it in the learning process~\cite{thomas2025llm, carless2018development}. To investigate the usefulness of the feedback, we therefore need to consider both what students receive and how they perform afterwards.

Handwritten mathematics poses an additional challenge compared with typed or structured answers. Before generating feedback, the model must interpret the notation, spatial structure, corrections, and order of the student's reasoning. Errors in this interpretation can carry into the diagnosis and feedback \cite{nath2025can, liu2026ai,kortemeyer2024grading}. Research in this area has mainly focused on recognition, error detection, and grading \cite{song2026can,sun2026circuit}. We extend this work by using handwritten university mathematics answers to generate individual feedback during a course. We assess the quality of this feedback and use a crossover design to examine whether receiving it improves later performance on exam questions testing the same mathematical skills.

We build on 
the pipeline detailed in~\cite{vanhoyweghen2026human}, in which handwritten assessments were scanned, graded with LLM support, and verified by human assessors. Once the grades were finalised, we used a separate generation step to provide comments on the student's answer and two practice exercises. The feedback was then sent by email, together with the grade and an image of the student's work. This allowed us to add individual feedback to the existing assessment process.

In this study, we describe how we implemented this approach and evaluate its use across six weekly bonus tests. We first assess a sample of the feedback using criteria that distinguish mathematical correctness from the quality of the diagnosis, guidance, and presentation. We then compare performance on related exam questions between students who did and did not receive feedback on the corresponding bonus test. Finally, we consider how attendance and the number of tests with feedback relate to overall exam performance.


\section{Methods}

\subsection{Participants and setting}

We collected data from the English- and Dutch-taught cohorts of the same undergraduate mathematics course at the Vrije Universiteit Brussel,
with 343 and 236 students with bonus-test records, respectively. Both cohorts shared many similarities; they were taught by the same professor and followed the same course content and structure. However, they also differed in important ways, including teaching assistants and backgrounds: the Dutch-speaking students came from the same schooling system, while the English-speaking students came from a wide range of backgrounds.

To investigate performance on related content, we linked specific bonus-test questions to final-exam questions at the start of the semester. Each cohort took a separate exam, but the topics and question types were closely matched. We linked the bonus-test data to exam records for 343 students in the English cohort and all 236 students in the Dutch cohort. Of these, 331 and 236 students participated in the first session exam.

\subsection{Bonus test design and A/B version groups}

Students could participate in six weekly bonus tests during the exercise sessions. Each test lasted 10 minutes, covered material from a previous exercise session, and consisted of two handwritten exercises. Participation was voluntary. The six tests could add up to 1.0 point to the final exam grade, which was graded out of 20.

Each test contained a common question (Q1) and a version-specific question (Q2). Within a class group, students assigned to versions A and B received the same Q1. We used this question as an anchor to assess comparability across version groups. Across class groups, the questions covered the same topic at a similar level of difficulty, but could differ in their wording or problem instance.

The version of Q2 depended on the parity of the student's identifier ($n_A^\text{ENG}=159$, $n_B^\text{ENG}=185$; $n_A^\text{NL}=121$, $n_B^\text{NL}=115$). Versions A and B covered the same topic but required different solution strategies. Consequently, when comparing performance on a linked exam question, we only include students who received the corresponding version of Q2.

\subsection{Grading and feedback pipeline}

\paragraph{Grading.}
We build on the grading pipeline set out in~\cite{vanhoyweghen2026human}. After scanning and anonymising the answer sheets, we used GPT-5.1 (Reasoning High) to score each response five times against a rubric-style grading key. We took the maximum score as the provisional grade, which a human grader then verified before release. We recorded all scores on a 0--10 scale.

\paragraph{Feedback randomisation.}
We varied feedback assignment on tests~1 and~3. On each of these tests, students were assigned to receive feedback on both questions or on neither question. The assignment followed a fixed crossover: students who received feedback on test~1 served as controls on test~3, and vice versa. We limited the randomised withholding of feedback to these two tests because extending it would have increased differences in the support students received on assessed course content. On the remaining tests, all attendees received feedback. Table~\ref{tab:design} reports the numbers in each group.

\begin{table}[ht]
\centering
\caption{Number of attendees assigned feedback or control on each bonus test. Assignment crossed over between tests~1 and~3. All attendees received feedback on the remaining tests. The duplicate Dutch record has been excluded.}
\label{tab:design}
\begin{tabular}{lcccc}
\toprule
& \multicolumn{2}{c}{\textbf{ENG}} & \multicolumn{2}{c}{\textbf{NL}} \\
\cmidrule(lr){2-3}\cmidrule(lr){4-5}
\textbf{Test} & Treated & Control & Treated & Control \\
\midrule
1 & 145 & 143 & 98 & 104 \\
2 & 277 & 0   & 172 & 0   \\
3 & 120 & 109 & 86 & 88  \\
4 & 235 & 0   & 146 & 0   \\
5 & 211 & 0   & 122 & 0   \\
6 & 196 & 0   & 123 & 0   \\
\bottomrule
\end{tabular}
\end{table}

\paragraph{Feedback generation.}
After verifying the grades, we used GPT-5.1 to generate feedback for each student's answer. We generated each feedback message only once. The model used the final grade to customize its comments and practice exercises to the student's performance, without altering the grade itself. For each question, we asked it to identify strengths and mistakes in the answer and provide two practice exercises.

To generate this feedback, we provided GPT-5.1 with the question text, a cropped image of the student's answer, the official worked solution, and the final numerical grade. No names or student identifiers were included. We used the following prompt for both cohorts:

\begin{tcolorbox}[breakable, colframe=mygrey, fontupper=\small, left=6pt, right=6pt, top=4pt, bottom=4pt]
You are supposed to give personalized feedback to a student's solution. Point out what they did well and what they could improve on. Be kind, constructive, and to the point. Use the solution provided to guide your feedback; you should not reference it explicitly. Treat the method used in the provided solution as the preferred method and keep your feedback and practice exercises in line with this method. You are not allowed to mention the final score. You can only use it to tailor your feedback.

In this course, we only work with real numbers. So if the student does not mention complex numbers, neither should you. Moreover, it's important to provide actionable feedback in the form of two sample exercises which they can use to improve their skills and practice more. For students with a score of 10, provide two similar exercises that are slightly more challenging. For students who did not attempt to solve the question, reassure them that it is not a problem and provide two simpler exercises to help them get started. For students who made a different exercise than the one asked, remind them to make the correct exercise in the future and give them two practice exercises. For all practice exercises, include only the final solution. Do not provide hints, explanations, or intermediate steps for these practice exercises. Students will not be able to respond to the feedback, or interact with you in any way. So please don't prompt them to interact with you.

\medskip
\textit{[Optional: Your feedback should be in Dutch.]}

\medskip
\textbf{Question:} \{question text\}\\
\textbf{Solution:} \{solution text\}\\
\textbf{Student's score:} \{student score\}\\
\textbf{Student answer:} \{anonymised answer image\}
\end{tcolorbox}

\paragraph{Post-processing and delivery.}
To ensure students could understand the feedback, we adapted the LLM-generated feedback, using \texttt{pylatexenc} and custom regular expressions to convert \LaTeX{} expressions and formatting into unformatted text. We also screened the generated feedback for potentially unintended or dangerous content and found none. Students received their final grade, an image of their answer, and, where assigned, the feedback. The email stated that the feedback was AI-generated and should be read critically. Students in the control group received only their grade. We delivered the feedback approximately one and a half week after each test.

\subsection{Evaluation of the generated feedback}

To evaluate the generated feedback, we sampled messages for three tasks from the English cohort: test~1, version~B, Q2; test~4, version~A, Q2; and test~5, Q1 in both versions. These tasks provided 82, 108, and 210 eligible messages, respectively. Each message concerned one student's answer to one question. We stratified the sample by the verified grade to include feedback for students at different performance levels. The score bands were $[0,1)$, $[1,4)$, $[4,7)$, $[7,10)$, and $\{10\}$, with target counts of 5, 15, 15, 15, and 5 per task. When a band contained fewer messages than the target, we included all of them. This resulted in 51, 51, and 55 messages from the three bonus tests.

Reviewers assessed each message alongside the student's answer, the question, the worked solution, and the verified grade. We used six criteria to distinguish mathematical accuracy, diagnostic specificity, actionability, method alignment, clarity and tone, and the quality of the practice exercises (Table~\ref{tab:rubric}). Each criterion was scored from 0 to~3, with 0 indicating deficient, 1 weak, 2 adequate, and 3 strong feedback. The rubric provided a separate description of each category for each criterion. Reviewers also completed a checklist indicating whether the feedback mentioned a strength, proposed an improvement or extension, gave a concrete next step, and included two practice exercises.

\begin{table}[htp]
\centering
\caption{Criteria and scoring guidelines used to evaluate the feedback. 
Scores were assigned relative to the student's answer and demonstrated level.}
\label{tab:rubric}
\footnotesize
\renewcommand{\arraystretch}{1.25}
\setlength{\tabcolsep}{3.5pt}
\begin{tabular}{
    @{}
    >{\raggedright\arraybackslash}p{3.2cm}
    >{\raggedright\arraybackslash}p{2.7cm}
    >{\raggedright\arraybackslash}p{2.7cm}
    >{\raggedright\arraybackslash}p{2.7cm}
    >{\raggedright\arraybackslash}p{2.7cm}
    @{}
}
\toprule
\textbf{Criterion} &
\textbf{0: Deficient} &
\textbf{1: Weak} &
\textbf{2: Adequate} &
\textbf{3: Strong} \\
\midrule

\textbf{A1: Mathematical accuracy}\newline
Correct evaluation of the student's work
&
Incorrect evaluation; invents errors or misses major ones
&
Partially correct evaluation with minor misjudgements
&
Correct overall, with minor gaps
&
Fully correct evaluation of the student's solution
\\

\addlinespace

\textbf{A2: Diagnostic specificity}\newline
Identification of strengths and weaknesses
&
No meaningful identification; generic or absent
&
Identifies the topic, but remains vague
&
Identifies specific elements that are correct or incorrect
&
Precisely identifies and explains why elements are correct or incorrect
\\

\addlinespace

\textbf{A3: Actionability}\newline
Usefulness of the guidance for further learning
&
Provides no guidance
&
Provides generic advice that the student cannot readily act on
&
Provides some concrete guidance, such as a correction or extension
&
Provides clear and targeted next steps for remediation or further learning
\\

\addlinespace

\textbf{A4: Method alignment}\newline
Consistency with the expected solution method
&
Conflicts with the expected method
&
Partially aligns with the expected method
&
Mostly aligns with the expected method
&
Fully aligns with the expected method
\\

\addlinespace

\textbf{A5: Clarity and tone}\newline
Quality of the communication
&
Confusing or inappropriate
&
Understandable, but poorly structured or verbose
&
Clear and reasonably concise
&
Clear, concise, and constructive
\\

\addlinespace

\textbf{A6: Practice-exercise quality}\newline
Relevance, correctness, and calibration of the exercises
&
Missing, incorrect, or irrelevant
&
Targets the wrong concept or level of difficulty
&
Relevant and correct, but weakly calibrated
&
Precisely targeted and well calibrated to the student's needs
\\

\bottomrule
\end{tabular}
\end{table}

\subsection{Linking bonus tests to exam questions}

At the start of the semester, we linked specific bonus-test items to final-exam questions covering the same mathematical skill. For example, version~B of Q2 in Test~1 concerned irrational inequalities, including domain restrictions and the conditions under which an inequality can be squared. The corresponding exam question assessed these same skills using a different inequality. Similarly, version~A of Q2 in Test~3 concerned limits at positive infinity involving fractions with square roots and polynomials. These links allowed us to examine later performance on the specific content addressed by the feedback.

The English- and Dutch-language exams were aligned and assessed the same types of mathematical questions, using different expressions. Table~\ref{tab:question_links} summarises the topics and methods linking the bonus-test items to the exam questions. For Q2, the relevant link depended on the student's assigned version.

\begin{table}[htbp]
\centering
\caption{Mathematical topics and methods linking bonus-test items to final-exam questions. Question references follow the English-language scoring file.}
\label{tab:question_links}
\small
\begingroup
\begin{tabular}{@{}p{3.1cm}p{1.8cm}
    >{\raggedright\arraybackslash}p{8.1cm}@{}}
\toprule
Bonus-test item & Exam item & Topic or method \\
\midrule
Test~1, Q2, B & Q6 &
Solving irrational inequalities involving square roots, with attention to domain and sign restrictions. \\[3pt]

Test~3, Q2, A & Q10A &
Evaluating limits at positive infinity involving square roots and polynomial growth. \\[3pt]

Test~4, Q2, A & Q7B &
Differentiating functions with a variable base and exponent, using logarithmic differentiation or an equivalent exponential form. \\[3pt]

Test~4, Q2, B & Q7C &
Applying the chain rule to composite functions with an outer square root. \\[3pt]

Test~5, Q1 & Q11A.ii &
Determining vertical and oblique asymptotes of rational functions with a cubic numerator and quadratic denominator. \\[3pt]

Test~5, Q2, A & Q11B.ii &
Determining intervals of convexity and concavity from the sign of a given second derivative. \\[3pt]

Test~5, Q2, B & Q11B.i &
Determining increasing and decreasing intervals from the sign of a given first derivative. \\[3pt]

Test~6, Q2, A & Q9A &
Evaluating indefinite integrals by substitution. \\[3pt]

Test~6, Q2, B & Q9B &
Evaluating indefinite integrals by integration by parts. \\
\bottomrule
\end{tabular}
\endgroup
\end{table}

\section{Results}

We first report participation and the evaluation of the feedback, followed by comparisons on linked exam questions and the descriptive results for overall exam performance.

\subsection{Sample characteristics}

Table~\ref{tab:desc} summarises the sample composition and participation patterns in the two cohorts. Students attended approximately four of the six bonus tests on average, with similar attendance across cohorts. Median final-exam performance was somewhat higher in the English cohort, which is likely a result of the fact that both courses were graded by different people, albeit with the same grading key. Nevertheless, the overall participation structure was comparable across the two language groups.

\begin{table}[htbp]
\centering
\caption{Sample characteristics by cohort.}
\label{tab:desc}
\begin{tabular}{lcc}
\toprule
 & \textbf{ENG} & \textbf{NL} \\
\midrule
Students with bonus-test records & 343 & {236} \\
Version A / Version B & {159 / 185} & {121 / 115} \\
{Numerical total exam scores} & 331 & {236} \\
Mean tests attended (SD) & {4.17 (1.79)} & 3.98 (1.81) \\
Exam Total/20: median (IQR) & 7.65 (3.91--11.33) & {6.38 (2.71--10.96)} \\
Crossover students ($n$) & 199 & 147 \\
$\rightarrow$ T1-treated / T3-treated & 95 / 104 & 75 / 72 \\
\bottomrule
\end{tabular}
\end{table}

\subsection{Evaluation of feedback quality}

\begin{figure}[htbp]
\centering
\includegraphics[width=\textwidth]{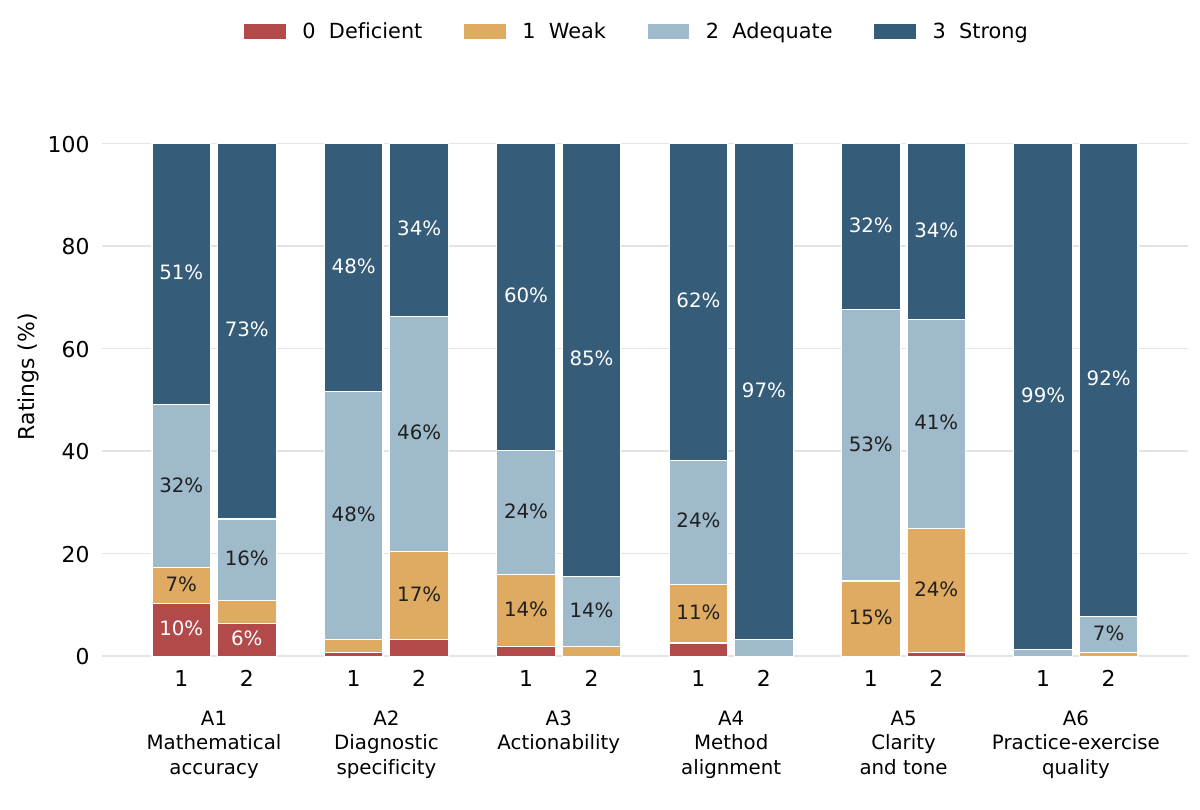}
\caption{Distribution of scores (0--3) by criterion. Each of the 157 messages was rated independently by two of the six reviewers; bars~1 and~2 show the first and second rating of each message (157 ratings per bar). The y-axis gives the percentage of ratings per score.}
\label{fig:rubric}
\end{figure}

Figure~\ref{fig:rubric} shows the distribution of ratings for each criterion and assignment. We find the highest ratings for the practice exercises (A6): 300 of the 314 ratings (95.5\%) were strong, with one weak rating and no deficient ratings. For mathematical accuracy (A1), 195 ratings (62.1\%) were strong, while 26 (8.3\%) were deficient. Thus, although the practice exercises were generally rated highly, the feedback did not always correctly evaluate the student's answer.

The proportion of strong ratings was lower for diagnostic specificity (A2) and clarity and tone (A5), at 129/314 (41.1\%) and 105/314 (33.4\%), respectively. Most of the remaining ratings were adequate. However, clarity and tone also received 61 weak ratings and one deficient rating, together accounting for 19.7\% of ratings. The comments give some indication of what reviewers found lacking. For example, feedback sometimes credited reasoning that was absent from the student's answer, missed an error, or suggested unnecessary improvements to a correct solution.

Referees differed the most for actionability and method alignment. For actionability, the first rating was strong for 131/155 messages (84.5\%), compared with 94/157 (59.9\%) for the second rating. For method alignment, these counts were 150/155 (96.8\%) and 97/157 (61.8\%), respectively. Moreover, we also observed lower actionability ratings for students who scored 0 or 1 out of 10. Within this group, the first rater assigned the highest rating to 16/28 retained messages (57.1\%), compared with 115/127 (90.6\%) for grades above 1/10. For the second rater, the corresponding proportions were 14/30 (46.7\%) and 80/127 (63.0\%). This difference may reflect that students who scored 0 or 1 on a question generally wrote very little, resulting in feedback that was more encouraging than specific. 

Alongside the feedback rubrics, we also included compliance checks. The model generally included the requested components. Reviewers recorded a strength in 933/942 reviews (99.0\%), an improvement or extension in 921/942 (97.8\%), concrete next steps in 902/942 (95.8\%), and two practice exercises in 932/942 (98.9\%).

\subsection{Crossover analysis on linked exam questions}

Table~\ref{tab:primary} reports the primary crossover comparisons for the two linked exam questions derived from tests~1 and~3. The estimated differences are mixed in direction across cohorts and questions, and at the level of individual linked exam questions the randomized evidence does not indicate a stable feedback effect.

This pattern is not entirely surprising in light of the broader feedback literature. Feedback effects are known to be sensitive to timing, specificity, and uptake \cite{kluger1996effects,shute2008focus}, and a single feedback exposure tied to one topic may simply be too light of an intervention to produce a reliable gain on one specific exam question asked months later.

\begin{table}[htbp]
\centering
\caption{Linked exam-question scores (out of 10) among students attending both crossover tests. Tests are two-sided Mann--Whitney comparisons, with Holm adjustment across the four rows.}
\label{tab:primary}
\small
\begin{tabular}{@{}llrrrrrr@{}}
\toprule
 & & \multicolumn{2}{c}{Feedback} & \multicolumn{2}{c}{Control} & & \\
\cmidrule(lr){3-4}\cmidrule(lr){5-6}
Cohort & Link & $n$ & Median & $n$ & Median & $p$ & $p_{\mathrm{Holm}}$ \\
\midrule
ENG & T1$\to$Q6   & 52 & 3.50 & 52 & 5.00 & 0.148 & 0.443 \\
ENG & T3$\to$Q10A & 51 & 2.00 & 41 & 2.00 & 0.670 & 0.670 \\
NL  & T1$\to$Q6   & 40 & 4.25 & 31 & 2.00 & 0.026 & 0.104 \\
NL  & T3$\to$Q10A & 40 & 1.75 & 33 & 5.00 & 0.230 & 0.461 \\
\bottomrule
\end{tabular}
\end{table}


\subsection{Descriptive patterns in attendance, feedback exposure, and exam performance}

\begin{figure}[htp]
\centering

\begin{subfigure}[t]{0.49\textwidth}
    \centering
    \includegraphics[width=\linewidth]{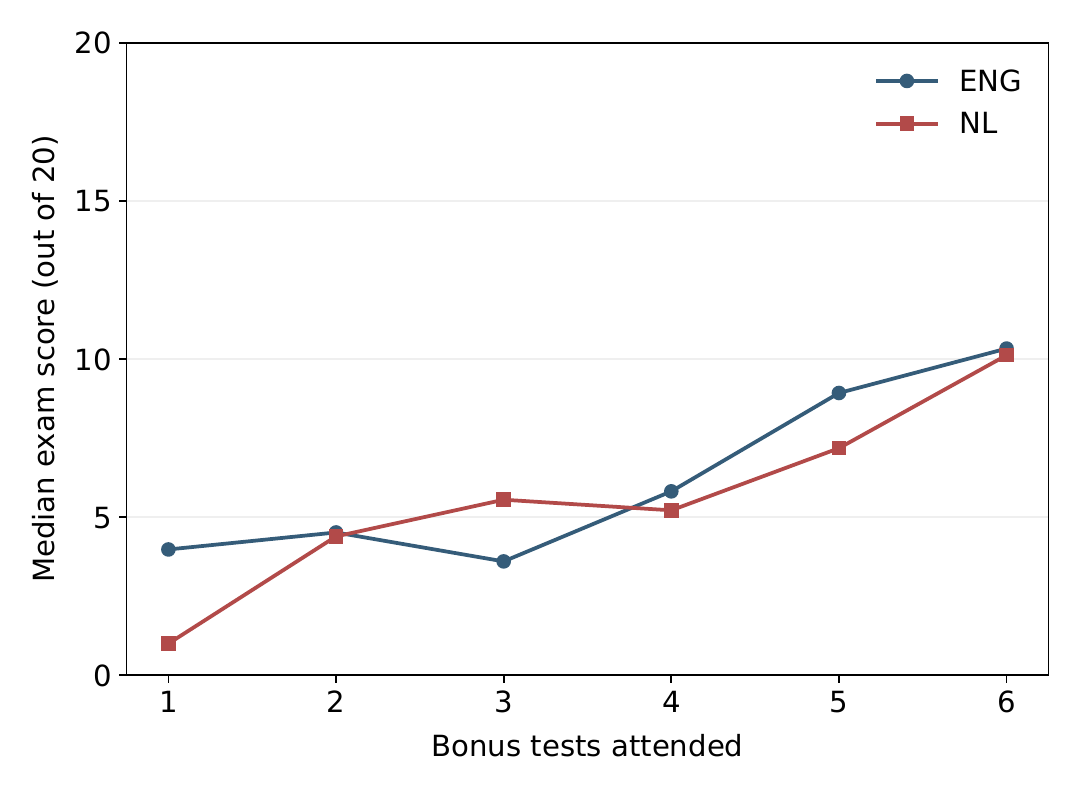}
    \caption{}
    \label{fig:dose}
\end{subfigure}
\hfill
\begin{subfigure}[t]{0.49\textwidth}
    \centering
    \includegraphics[width=\linewidth]{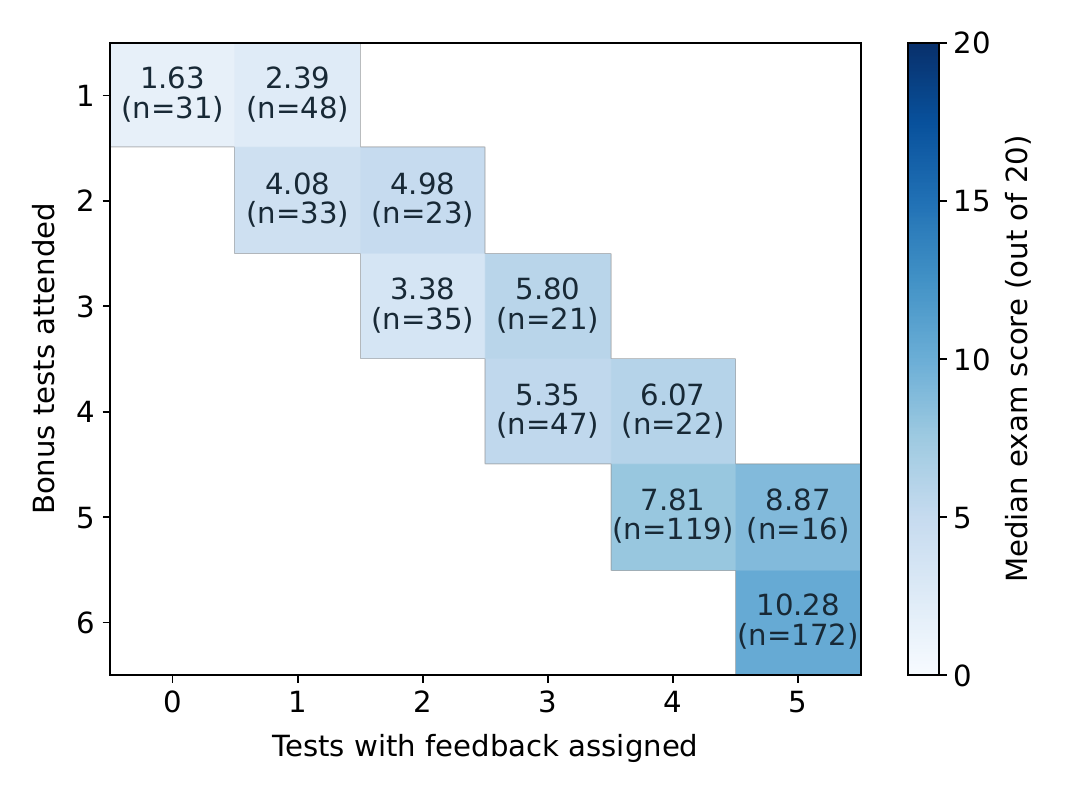}
    \caption{}
    \label{fig:heatmap}
\end{subfigure}

\caption{\textbf{(a)} Median exam score out of 20, before bonus points, by the number of bonus tests attended, showing an attendance effect in both cohorts. \textbf{(b)} Median exam score, before bonus points, as a joint function of bonus-test attendance and cumulative exposure to AI-generated feedback. Within all attendance levels, greater feedback exposure is associated with higher median exam performance.}
\label{fig:attendance_feedback}
\end{figure}

When considering overall exam performance, we observe a positive association with attendance in both cohorts (Figure~\ref{fig:dose}; Spearman $\rho_{\text{ENG}}=0.49$, $\rho_{\text{NL}}=0.54$, both $p<0.001$). Students who attended more bonus tests tended to obtain higher exam scores. However, they also had more opportunities to practise and may have differed in their engagement with the course. 

Furthermore, figure~\ref{fig:heatmap} compares the median exam performance within attendance levels as a joint function of attendance and cumulative feedback exposure. Within all attendance strata, students with more feedback exposure also achieved higher median exam scores. A simple descriptive summary of these within-attendance contrasts suggests a conditional mean advantage of about 0.5 grade points and a conditional median advantage of about 0.9 grade points for the more-exposed groups. Because feedback exposure outside the crossover tests was driven by which bonus tests students attended, these within-attendance contrasts cannot be cleanly attributed to feedback alone. Even so, the pattern is consistent with the possibility that repeated AI-generated feedback could support learning gains across the semester, rather than only performance on a single later exam question.

\section{Discussion}
In this study, we examined both the quality of AI-generated feedback on handwritten mathematics and its contribution to subsequent exam performance. The clearest positive result concerned the practice exercises: 95.5\% of ratings placed them in the highest category for relevance, correctness, and calibration to the student's needs. The feedback itself received more mixed assessments, particularly for diagnostic specificity and clarity. These findings show that the model could provide suitable additional practice even when its explanation of the student's answer needed improvement. Informal comments collected from students attending the final lecture were consistent with these ratings. Several students reported that the feedback helped them identify what to work on or recognise what they had done well. Others described explanations as unclear or superficial, pointed out errors, or objected to the use of AI. Some also expressed disappointment when feedback was absent. Although these comments come from a self-selected group and cannot be taken as representative of either cohort, they point to the same weaknesses as the reviewer ratings.

The lower ratings for diagnostic specificity and clarity echo previous trepidations with respect to AI feedback~\cite{scholz2025partnering}. Reviewers described feedback that attributed reasoning to students which they had not used, proposed unnecessary corrections, or gave overly extensive explanations. These observations are relevant to the prompt, which required the model to identify both strengths and areas for improvement. When a student has answered correctly, this instruction may encourage the model to look for a weakness where little needs to be changed. For students who scored 0 or 1, the problem can be different. As R1 explained, a correct account of how to solve the entire exercise may still leave the student unsure how to proceed. We believe this distinction between correctness and usefulness deserves particular attention when generating feedback for students who struggle with the material. The practice exercises in our setup included only final answers, as the prompt explicitly excluded hints and intermediate steps. Providing a worked example or a first step could help these students begin practising. These observations suggest that the guidance should depend on how much progress the student has made. Students who have barely started may need help with a first step, while those with a correct answer may benefit more from an extension than from suggestions for improvement. Whether adapting the prompt in this way improves the feedback and subsequent performance remains to be tested.

The positive relation between attendance, feedback exposure, and overall exam scores may suggest a direct benefit from receiving AI feedback. However, students who attended more tests also practised more, while differences in exposure at a given attendance level depended on which tests they attended. These observations therefore cannot resolve the inconclusive results of the randomised comparisons. We would need a design that varies repeated feedback exposure more extensively to assess whether its effects accumulate across the semester.

Several limitations should be considered when interpreting our findings. Each feedback rubric was graded by only two referees, who mostly aligned, but important differences existed, especially in the specificity and actionability of the feedback. Within the exam analysis, the comparison groups were relatively small and restricted to students who attended both crossover tests and had an observed outcome. Conditioning on attendance may affect the comparability created by random assignment, and feedback on one topic could also directly affect performance on another subject; for instance, feedback on limits could impact students' ability to recognise oblique asymptotes. Students were informed that the generated feedback was AI-generated, which in turn could have lowered their confidence or reliance on the provided feedback \cite{nazaretsky2026gives, henderson2026comparing}; some informal student comments indicated objections to the use of AI.

Our findings show that GPT-5.1 can generate practice exercises from handwritten mathematics answers that reviewers judge relevant, correct, and suited to the student's level. They also identify where the accompanying feedback needs improvement, particularly in diagnosing mistakes and explaining how to proceed. The crossover comparisons did not establish statistically significant improvements in later exam performance. The positive associations with repeated feedback exposure warrant further investigation in a design that further separates feedback from attendance and practice. Recording whether students read the feedback and attempt the exercises would also help explain when the assessed quality of the material translates into learning.

\section*{Declaration of Conflicting Interest} 
\addcontentsline{toc}{section}{Declaration of Conflicting Interest} 
The authors declare that they have no known competing financial interests or personal relationships that could have appeared to influence the work reported in this paper.

\section*{Funding}
\addcontentsline{toc}{section}{Funding}

This research was supported by funding from the Vrije
Universiteit Brussel Research Council (VUB-OZR) and the Flemish Government under the ``Onderzoeksprogramma Artifici\"ele Intelligentie (AI) Vlaanderen'' program.
Andres Algaba acknowledges support from the Francqui Foundation (Belgium) through a Francqui Start-Up Grant and a fellowship from the Research Foundation Flanders (FWO) under Grant No.1286924N. 
Vincent Ginis acknowledges support from Research Foundation Flanders under Grant No.G032822N and G0K9322N. 

\subsection*{Data Availability}
The computer data has been made available online at:~\url{https://figshare.com/s/9208da7e9bc80b2cd01f}.

\subsection*{Declaration of generative AI and AI-assisted technologies in the writing process}
\addcontentsline{toc}{section}{Declaration of generative AI and AI-assisted technologies in the writing process} 
During the preparation of this work, the authors used ChatGPT-5.6 to improve the manuscript's readability. After using this tool/service, the authors reviewed and edited the content as needed and take full responsibility for the content of the publication.

\subsection*{Ethics approval and consent to participate}
\addcontentsline{toc}{section}{Ethics approval and consent to participate} 
This study retrospectively analyses pseudonymised educational assessment data generated through routine teaching practice from an existing formative assessment structure embedded in the course curriculum. Under institutional guidance on the reuse of educational/secondary data, no separate ethical review was required.

\phantomsection

\bibliographystyle{unsrt}
\bibliography{references} 

@article{black1998assessment,
  title={Assessment and classroom learning},
  author={Black, Paul and Wiliam, Dylan},
  journal={Assessment in Education: principles, policy \& practice},
  volume={5},
  number={1},
  pages={7--74},
  year={1998},
  publisher={Taylor \& Francis}
}

@article{hattie2007power,
  author    = {Hattie, John and Timperley, Helen},
  title     = {The power of feedback},
  journal   = {Review of Educational Research},
  year      = {2007},
  volume    = {77},
  number    = {1},
  pages     = {81--112},
  doi       = {10.3102/003465430298487}
}

@article{kasneci2023chatgpt,
  title={ChatGPT for good? On opportunities and challenges of large language models for education},
  author={Kasneci, Enkelejda and Se{\ss}ler, Kathrin and K{\"u}chemann, Stefan and Bannert, Maria and Dementieva, Daryna and Fischer, Frank and Gasser, Urs and Groh, Georg and G{\"u}nnemann, Stephan and H{\"u}llermeier, Eyke and others},
  journal={Learning and individual differences},
  volume={103},
  pages={102274},
  year={2023},
  publisher={Elsevier}
}

@article{kluger1996effects,
  title={The effects of feedback interventions on performance: a historical review, a meta-analysis, and a preliminary feedback intervention theory.},
  author={Kluger, Avraham N and DeNisi, Angelo},
  journal={Psychological bulletin},
  volume={119},
  number={2},
  pages={254},
  year={1996},
  publisher={American Psychological Association}
}

@article{mollick2023assigning,
  title={Assigning AI: Seven approaches for students, with prompts},
  author={Mollick, Ethan and Mollick, Lilach},
  journal={arXiv preprint arXiv:2306.10052},
  year={2023}
}

@article{nicol2006formative,
  title={Formative assessment and self-regulated learning: A model and seven principles of good feedback practice},
  author={Nicol, David J and Macfarlane-Dick, Debra},
  journal={Studies in higher education},
  volume={31},
  number={2},
  pages={199--218},
  year={2006},
  publisher={Taylor \& Francis}
}

@article{shute2008focus,
  title={Focus on formative feedback},
  author={Shute, Valerie J},
  journal={Review of educational research},
  volume={78},
  number={1},
  pages={153--189},
  year={2008},
  publisher={Sage Publications}
}

@article{vanhoyweghen2026human,
  title={Human-in-the-Loop LLM Grading for Handwritten Mathematics Assessments},
  author={Vanhoyweghen, Arne and Holst, Vincent and Mobini, Melika and Van de Voorde, Lukas and Vanleke, Tibo and Verbruggen, Bert and Verbeken, Brecht and Algaba, Andres and Verboven, Sam and Guerry, Marie-Anne and others},
  journal={arXiv preprint arXiv:2603.13083},
  year={2026}
}

@article{wisniewski2020power,
  title={The power of feedback revisited: A meta-analysis of educational feedback research},
  author={Wisniewski, Benedikt and Zierer, Klaus and Hattie, John},
  journal={Frontiers in psychology},
  volume={10},
  pages={3087},
  year={2020},
  publisher={Frontiers Media SA}
}

@inproceedings{meyer2026comparing,
  title={Comparing Teacher and AI-Generated Feedback in the Writing Classroom: Experimental Results from Secondary School Classrooms},
  author={Meyer, Jennifer and Steinbach, Marlene and Schiller, Ronja and Mertens, Ute and Schaller, Nils-Jonathan and Horbach, Andrea and Fleckenstein, Johanna and K{\"o}ller, Olaf and Kizilcec, Rene F and Jansen, Thorben},
  booktitle={Proceedings of the Thirteenth ACM Conference on Learning@ Scale},
  pages={264--276},
  year={2026}
}

@article{dai2024assessing,
  title={Assessing the proficiency of large language models in automatic feedback generation: An evaluation study},
  author={Dai, Wei and Tsai, Yi-Shan and Lin, Jionghao and Aldino, Ahmad and Jin, Hua and Li, Tongguang and Ga{\v{s}}evi{\'c}, Dragan and Chen, Guanliang},
  journal={Computers and Education: Artificial Intelligence},
  volume={7},
  pages={100299},
  year={2024},
  publisher={Elsevier}
}

@inproceedings{scholz2025partnering,
  title={Partnering with AI: a pedagogical feedback system for LLM integration into programming education},
  author={Scholz, Niklas and Nguyen, Manh Hung and Singla, Adish and Nagashima, Tomohiro},
  booktitle={European Conference on Technology Enhanced Learning},
  pages={243--248},
  year={2025},
  organization={Springer}
}

@inproceedings{thomas2025llm,
  title={LLM-generated feedback supports learning if learners choose to use it},
  author={Thomas, Danielle R and Borchers, Conrad and Bhushan, Shambhavi and Gatz, Erin and Gupta, Shivang and Koedinger, Kenneth R},
  booktitle={European Conference on Technology Enhanced Learning},
  pages={489--503},
  year={2025},
  organization={Springer}
}

@inproceedings{steinbach2025llms,
  title={When LLMs hallucinate: Examining the effects of erroneous feedback in math tutoring systems},
  author={Steinbach, Marlene and Bhandari, Shreya and Meyer, Jennifer and Pardos, Zachary A},
  booktitle={Proceedings of the Twelfth ACM Conference on Learning@ Scale},
  pages={139--150},
  year={2025}
}

@article{baral2024automated,
  title={Automated feedback in math education: A comparative analysis of LLMs for open-ended responses},
  author={Baral, Sami and Worden, Eamon and Lim, Wen-Chiang and Luo, Zhuang and Santorelli, Christopher and Gurung, Ashish and Heffernan, Neil},
  journal={arXiv preprint arXiv:2411.08910},
  year={2024}
}

@article{nazaretsky2026gives,
  title={Who Gives Feedback Matters: Student biases towards human and AI-generated formative feedback},
  author={Nazaretsky, Tanya and Mejia-Domenzain, Paola and Swamy, Vinitra and Frej, Jibril and K{\"a}ser, Tanja},
  journal={Journal of Computer Assisted Learning},
  volume={42},
  number={1},
  pages={e70153},
  year={2026},
  publisher={Wiley Online Library}
}

@article{henderson2026comparing,
  title={Comparing Generative AI and teacher feedback: student perceptions of usefulness and trustworthiness},
  author={Henderson, Michael and Bearman, Margaret and Chung, Jennifer and Fawns, Tim and Buckingham Shum, Simon and Matthews, Kelly E and de Mello Heredia, Jimena},
  journal={Assessment \& Evaluation in Higher Education},
  volume={51},
  number={5},
  pages={863--878},
  year={2026},
  publisher={Taylor \& Francis}
}

@article{liu2026ai,
  title={AI-assisted automated short answer grading of handwritten university-level mathematics exam},
  author={Liu, Tianyi and Chatain, Julia and Kobel-Keller, Laura and Kortemeyer, Gerd and Willwacher, Thomas and Sachan, Mrinmaya},
  journal={Teaching Mathematics and Its Applications},
  volume={45},
  number={1},
  pages={84--105},
  year={2026},
  publisher={Oxford University Press}
}

@inproceedings{nath2025can,
  title={Can Vision-Language Models Evaluate Handwritten Math?},
  author={Nath, Oikantik and Bathina, Hanani and Khan, Mohammed Safi Ur Rahman and Khapra, Mitesh M},
  booktitle={Proceedings of the 63rd Annual Meeting of the Association for Computational Linguistics (Volume 1: Long Papers)},
  pages={14784--14814},
  year={2025}
}

@inproceedings{song2026can,
  title={Can MLLMs Read Students' Minds? Unpacking Multimodal Error Analysis in Handwritten Math},
  author={Song, Dingjie and Xu, Tianlong and Zhang, Yi-Fan and Li, Hang and Yan, Zhiling and Fan, Xing and Li, Haoyang and Sun, Lichao and Wen, Qingsong},
  booktitle={International Conference on Artificial Intelligence in Education},
  pages={110--124},
  year={2026},
  organization={Springer}
}

@inproceedings{sun2026circuit,
  title={EDU-CIRCUIT-HW: Evaluating Multimodal Large Language Models on Real-World University-Level STEM Student Handwritten Solutions},
  author={Sun, Weiyu and Chen, Liangliang and Cai, Yongnuo and Xie, Huiru and Zeng, Yi and Zhang, Ying},
  booktitle={Findings of the Association for Computational Linguistics: ACL 2026},
  pages={15281--15314},
  year={2026}
}

@article{carless2018development,
  title={The development of student feedback literacy: Enabling uptake of feedback},
  author={Carless, David and Boud, David},
  journal={Assessment \& evaluation in higher education},
  volume={43},
  number={8},
  pages={1315--1325},
  year={2018},
  publisher={Taylor \& Francis}
}

@article{pardo2019using,
  title={Using learning analytics to scale the provision of personalised feedback},
  author={Pardo, Abelardo and Jovanovic, Jelena and Dawson, Shane and Ga{\v{s}}evi{\'c}, Dragan and Mirriahi, Negin},
  journal={British journal of educational technology},
  volume={50},
  number={1},
  pages={128--138},
  year={2019},
  publisher={Wiley Online Library}
}

@article{kortemeyer2024grading,
  title={Grading assistance for a handwritten thermodynamics exam using artificial intelligence: An exploratory study},
  author={Kortemeyer, Gerd and N{\"o}hl, Julian and Onishchuk, Daria},
  journal={Physical Review Physics Education Research},
  volume={20},
  number={2},
  pages={020144},
  year={2024},
  publisher={APS}
}


\end{document}